\documentclass[letterpaper]{JHEP3}
\usepackage{epsfig,verbatim}
\usepackage{subfigure}
\usepackage{amsmath, amssymb, graphics}
\usepackage{slashed}            
\usepackage{bbm}                

\newcommand{\mathsym}[1]{{}}

\newcommand{\eref}[1]{(\ref{#1})}

\renewcommand\({\left(}
\renewcommand\){\right)}
\renewcommand\[{\left[}
\renewcommand\]{\right]}

\newcommand{\dd}{{\rm d}}
\newcommand{\e}{{\rm e}}

\newcommand\eps{\epsilon}

\newcommand\mn{{\mu\nu}}

\def\ba{\begin{eqnarray}}
\def\ea{\end{eqnarray}}
\def\be{\begin{equation}}
\def\ee{\end{equation}}

\def\L{\mathcal{L}}
\def\O{\mathcal{O}}

\def\D{\mathcal{D}}

\def\H{\mathcal{H}}

\def\nn{\nonumber}
\def\({\left(}
\def\){\right)}

\def\eref#1{(\ref{#1})}

\newcommand{\roughly}[1]{\mathrel{\raise.3ex\hbox{$#1$\kern-0.85em
\lower1ex\hbox{$\sim$}}}}

\title{Goldstone bosons and a dynamical Higgs field}

\author{Sander Mooij\footnote{smooij@nikhef.nl}  $\,$and Marieke
Postma\footnote{mpostma@nikhef.nl} \\ Nikhef, Science Park 105, 1098
XG Amsterdam, The Netherlands.}

\date{}

\abstract {Higgs inflation uses the gauge variant Higgs field as the
  inflaton. During inflation the Higgs field is displaced from its
  minimum, which results in associated Goldstone bosons that are apparently massive. Working in a minimally coupled $U(1)$ toy model, we use the closed-time-path formalism to show that
  these Goldstone bosons do contribute to the one-loop effective
  action. Therefore the computation in unitary gauge gives incorrect
  results. Our expression for the effective action is gauge invariant
  upon using the background equations of motion. }

\preprint{NIKHEF 2011-010}

\begin{document}

\section{Introduction}

The mechanism of Higgs inflation is already an old idea
\cite{Salopek:1988qh}, which was recently revived by Bezrukov and
Shaposhnikov \cite{Bezrukov:2007ep,bezrukov2,bezrukov3}. It is elegant
in its simplicity: why look for exotic inflatons if the Standard Model
already possesses a viable candidate?  Inflation is obtained by
introducing an additional coupling between the Higgs field and the
Ricci scalar $\mathcal{R}$.  It offers the exciting possibility that
the Higgs mass can be predicted from cosmological data on the cosmic
microwave background (CMB) \cite{popa,kiselev}.  This requires the
computation of the quantum corrections to the potential
\cite{starobinsky1, starobinsky2, barvinsky2,wilczek,kaiser,hertzberg,cliff,cristiano,lerner1,lerner2,bezrukov4,lerner3,calmet,barbon}.
In this work we want to clarify the role played by Goldstone bosons in
the loop calculation.

During inflation and the reheating period afterwards, the Higgs field
is evolving in its potential.  This complicates the calculation of the
one-loop effective action compared to the vacuum calculation, with the
Higgs field in the minimum of its potential. First of all, the
Goldstone bosons are now apparently massive and, as we will show, do
contribute to the one-loop effective action (see also
\cite{starobinsky1, starobinsky2}). Second, there are time-dependent
corrections to the Coleman-Weinberg expression which strictly only
applies to the static case.  Both effects have not been fully
appreciated in the literature as they are small during inflation.
However, they are important afterwards, and should be taken into
account if one wants to relate low energy observables (the Higgs mass
to be measured at the LHC) to high energy (CMB) observables.

Goldstone's theorem states that there is one massless boson for each
generator of a continuous symmetry that is broken spontaneously by the
ground state. In a gauge theory these Goldstone bosons do not appear
as independent physical particles. They are ``eaten''  by the gauge
bosons; their associated degree of freedom (d.o.f.) is used to turn a
massless vector boson (2 d.o.f.)  into a massive one (3 d.o.f.). This
is best seen in unitary gauge, in which the Goldstone bosons
explicitly disappear from the theory.

During the cosmological evolution of the Higgs field this picture
changes. The Higgs field is displaced from its minimum, and is
evolving in time.  The gauge symmetry is broken, but the associated
Goldstone bosons are no longer massless eigenstates. They can still
be removed from the theory by going to unitary gauge, (though only upon
using the equations of motion). Therefore one might be inclined to
think that the Goldstone bosons are still unphysical, and that their
contribution to any quantum corrections should be omitted. This would
be dramatic for supersymmetric Higgs inflation
\cite{einhorn,ferrara1,ferrara2}, as the quadratic corrections would
no longer cancel.

Potential problems with calculating quantum corrections in unitary
gauge were noted before in the literature \cite{jackiw,rachel}. To
investigate the effect of the massive Goldstone bosons, we use the
closed-time-path formalism
\cite{ctpschwinger,ctpkeldysh,Bakshi1,Bakshi2,ctpjordan,ctpcalzetta,ctppaz,ctpweinberg}
to compute Coleman-Weinberg one-loop corrections. In this work we
restrict ourselves to a minimally coupled U(1) toy model in flat
spacetime. {\it We find that corrections induced by the U(1) Goldstone
boson are real and can not be omitted.}  Our results apply to Standard
Model Higgs inflation, as well as to models in which the inflaton is a
Higgs field of some grand unified theory \cite{rachel2}
\cite{Antusch:2010va}. In addition, we calculate the corrections due
to the time-dependence of the Higgs field.  These are essential for
showing that our result is gauge invariant.

A large part of our computation follows the work by Heitmann and
Baacke
\cite{baacke1,baacke2,baacke3,baacke4,heitmann,Heitmannmt,Heitmannphdt}. We
generalize their results for an arbitrary Higgs potential.  We
calculate the equation of motion for the background field rather than
the effective action directly; up to a field-independent constant the
latter can always be obtained by formally integrating the field equations.  Our
results reduce to the original Coleman-Weinberg result in the static
limit. Our calculation is done in $R_\xi$ gauge.  Boyanovsky et
al. have calculated the one-loop potential in terms of gauge invariant
quantities \cite{boyanovski}, but only in the adiabatic limit, which
does not take into account the time-dependence of the rolling Higgs
field.

We will be working in Minkowski spacetime, with $\{+---\}$ signature,
and set $\hbar=c=k_B=1$. We choose Feynman-'t Hooft gauge $\xi =1$.
In the appendix we calculate the equation of motion perturbatively, in
arbitrary $R_\xi$ gauge. There we show that the gauge-dependent terms
cancel upon using the equation of motion for the background field
$\phi$. The effective potential has already been shown to be gauge
invariant when calculated around a potential minimum
\cite{jackiw,kugo}. Here we show that gauge invariance holds
also in this more general case at the one-loop level, but only
on-shell, upon using the background equations of motion.

The article is organized as follows. In the next section we discuss
the Abelian Higgs model at the classical level.  We start by
generalizing Goldstone's theorem to the case with the Higgs field
displaced from its minimum, relating the Goldstone boson mass to the
slope of the Higgs potential. Although apparently massive, the
Goldstone bosons can still be removed from the theory in unitary
gauge, but only upon using the equations of motion. We end the section
with a discussion of the problems encountered if one attempts to
calculate the one-loop effective action in unitary gauge
\cite{jackiw,rachel}.  To resolve these problems we calculate the
equations of motion in section \ref{berekening}. The calculation is
set up in a non-perturbative way.  However, to extract the divergent
parts explicitly, we use a perturbative expansion.  We end with a
discussion of our results in section \ref{conclusions}.  A brief
outline of the CTP formalism, and our definitions and conventions
used, are relegated to appendix \ref{ctp}. In appendix \ref{pert} we
present a perturbative calculation of the equations of motion in
arbitrary $R_\xi$ gauge. Although more technically involved, it shows
explicitly that the results are gauge invariant upon using the
background equation of motion.

\section{The rolling Goldstone boson}

In this section we show how the usual Goldstone boson theorem
\cite{Goldstone1,Goldstone2} changes when we consider a global U(1)
symmetry broken by a scalar field that is {\it not} in its minimum.
We then discuss how this affects the Higgs mechanism in the gauged
version of the theory.  It still seems possible to go to unitary
gauge. However, studying the associated Coleman-Weinberg corrections
suggests a problem with this gauge.  For simplicity, we will focus on
a U(1) gauge theory. The results can be easily generalized to
non-Abelian gauge groups.

\subsection{Goldstone boson theorem}

Consider a theory with a complex scalar field $\Phi$, which we will
refer to as the Higgs field. It is invariant under a global U(1)
transformation.  The field has a time-dependent expectation value
$\Phi_{\rm cl} = (\phi_R(t) + i \phi_I(t))/\sqrt{2}$; without loss of
generality we can align this with the real direction and set $\phi_I
=0$. Goldstone showed that in the broken phase $\phi_R \neq 0$ there
is a massless excitation in the spectrum, provided the potential is
extremized \cite{Goldstone1,Goldstone2}.  Here we repeat his argument
for a (time-dependent) classical background field which is displaced
from its minimum $\partial_{\phi_R} V|_{\rm cl} \neq 0$.

Under an infinitesimal global U(1) transformation $\Phi \to
\e^{i\alpha} \Phi$ the invariant potential $V(\Phi \Phi^\dagger)$
transforms as
\be \delta_\alpha V = \frac{\partial V}{\partial {\phi_i}}
\delta_\alpha \phi_i =0,
\label{gaugeV}
\ee
with $i = \{R,I\}$. Written out in terms of real fields the change
under a gauge transformation is $\delta_\alpha \phi_R = - \alpha
\phi_I$ and $\delta_\alpha \phi_I = \alpha \phi_R$.  Differentiating
\eref{gaugeV} with respect to $\phi_k$, the equation for $k=R$ is
trivially satisfied. For $k=I$ evaluated on the classical background
configuration it yields, however,
\be
\frac{\partial^2 V}{\partial {\phi_I}\partial {\phi_I}} \phi_R -
\frac{\partial V}{\partial {\phi_R}}\bigg|_{\rm cl} =0.
\label{goldstone}
\ee 
If the Higgs extremizes the potential, the second term in the equation
above vanishes. One concludes that the spectrum contains a massless
Goldstone boson.  However, with the Higgs displaced from its minimum
--- as is the case during Higgs inflation --- the first derivative of
the potential no longer vanishes. Therefore the Goldstone boson mass
is apparently non-zero:
\be m_I^2  \equiv \frac{\partial^2 V}{\partial \phi_I^2} \bigg|_{\rm cl}
= \frac{1}{\phi_R} \frac{\partial V}{\partial \phi_R} \bigg|_{\rm cl}
= - \frac{\ddot \phi_R}{\phi_R}\bigg|_{\rm cl}. 
\label{mI} 
\ee 
Strictly speaking, we can only unambiguously identify the mass of
excited states with the (eigenstates of the) second derivative of the
potential if the potential is minimized. In a time-dependent
background fields may mix non-trivially in the kinetic terms as well.
Throughout the paper we will be sloppy with this distinction and
equally use ``mass matrix'' and ``second derivative of the potential''
$m_{ij} \equiv V_{\phi_i \phi_j}$, as was done in \eref{mI} above.
The last equality is only valid on-shell, as we used that the
evolution of the classical background $\phi_R(t)$ is governed by the
Klein-Gordon equation, which in a Minkowski universe reads $\ddot
\phi_R + \partial_{\phi_R} V =0$.

\subsection{Higgs mechanism} \label{hm}

We now gauge the U(1) model of the previous section. How does the Higgs
mechanism work during inflation, when the Higgs is displaced from
its minimum and the Goldstone boson is massive?  The standard
lore found in textbooks is that the gauge boson cannot obtain a mass,
unless this mass term is associated with a pole in the vacuum
polarization amplitude, which can only be created by a massless scalar
particle.

The Lagrangian of the U(1) Abelian Higgs model is
\be
\L = - \frac14 F_{\mu \nu}F^{\mu \nu} + D_\mu \Phi (D^\mu \Phi)^\dagger - 
V(\Phi \Phi^\dagger),
\label{L_abelian}
\ee
with $F_{\mu\nu}$ the Abelian field strength and $D_\mu \Phi =
(\partial_\mu + ig A_\mu)\Phi$ the covariant derivative.  Under a
$U(1)$ gauge transformation the Higgs and gauge field transform
\be \Phi \to \e^{i \alpha} \Phi, \qquad A_\mu \to A_\mu -\frac{1}{g}
\partial_\mu \alpha, 
\label{gaugetrafo}
\ee
with $\alpha$ the infinitesimal parameter of the gauge transformation,
and $g$ the U(1) gauge coupling.
To analyze the Higgs
mechanism we perturb the Higgs field around the classical background:
\ba \Phi (x,t) 
&=& \frac1{\sqrt{2}} \( \Phi_R(x,t) + i\Phi_I(x,t)\) 
=\frac1{\sqrt{2}} \big[
(\phi_R(t) + h(x,t)) + i \theta(x,t) \big], \nn\\
A_\mu(x,t)  &=& A_\mu(x,t),
\label{flds_exp}
\ea
with as before $\phi_R(t)$ the classical background field, and
$h(x,t)$, $\theta(x,t)$, $A_\mu(x,t)$ the fluctuations of the Higgs
and gauge field respectively.

The potential $V(\Phi \Phi^\dagger)$ can be expanded in the perturbed
fields
\be V = V|_{\rm cl} +  V_R|_{\rm cl} \, h
+ \frac 12 V_{RR} |_{\rm cl} \, h^2
+ \frac 12 V_{II} |_{\rm cl} \, \theta^2
+ ...
\label{expV}
\ee
with the dots representing terms of cubic order or higher in the
fluctuations. Here we introduced the notation $V_i = \partial_{\Phi_i}
V$.  Because of the U(1) symmetric form of the potential there are no
terms linear in $\theta$.  There is however a tadpole term in $h$ if
the Higgs is displaced from its minimum.  Similarly we expand the
kinetic terms:
\ba
\L_{\rm kin} 
&=&-\frac14  F_{\mu \nu}F^{\mu \nu}
+ \frac12 \( \partial_\mu h \partial^\mu h
+ \partial_\mu \theta \partial^\mu \theta + g^2 \phi_R^2  A_\mu A^\mu \) +
g \phi_R A_\mu \partial^\mu \theta 
\nn \\ && - g\dot \phi_R \theta A_0 + \dot h \dot \phi_R
+ \frac12 \dot \phi_R^2 + \dots
\label{expK}
\ea
The terms in the second line are absent for a Higgs field in a static
minimum.

Now we transform to unitary gauge. Define a new gauge field via
\be
A_\mu = B_\mu - \frac1{g} \partial_\mu(\theta/\phi_R).
\label{AtoB}
\ee
This leaves the potential and the kinetic term for the gauge fields
invariant, but affects the Higgs kinetic terms. Writing the kinetic
Lagrangian in terms of the newly defined field $B_\mu$ removes the
kinetic term for the Goldstone $\theta$ and its derivative
coupling to the gauge field:
\be \L_{\rm kin} = -\frac14 B_{\mu \nu} B^{\mu \nu}
+ \frac12 \( \partial_\mu h \partial^\mu h + g^2 \phi_R^2 B_\mu B^\mu \) 
-\frac{\theta^2 \dot \phi_R^2}{2 \phi_R^2} +
\frac{\theta \dot \theta \dot \phi_R}{\phi_R} + \dot h \dot \phi_R +
\frac12 \dot \phi_R^2 +\dots 
\label{LkinB}
\ee
where $B_{\mu\nu}$ is the Abelian field strength for $B_\mu$. If the
potential is minimized we have $V_R =0$ and $\dot \phi_R =0$, as in
the usual description of the Higgs mechanism. The Goldstone boson
completely disappears from the Lagrangian. It is eaten by the
longitudinal component of the gauge field $A_L$ which has become
massive: $m_A = g \phi_R$.  However, with the Higgs displaced from its
minimum, the Goldstone boson cannot be eliminated from the Lagrangian
by the field redefinition \eref{AtoB}, or equivalently by a unitary
gauge transformation \eref{gaugetrafo} with $\alpha = \theta/\phi_R$.
The $\theta$-field is still present, both in the kinetic and in the
potential part of the Lagrangian.  Nevertheless, the gauge field has
still become massive. How is this possible without a massless pole in
the polarization tensor? The answer lies in the last four
time-dependent terms in \eref{LkinB}. These exactly cancel the
Goldstone mass term in \eref{expV} when the fields are taken on-shell.
Indeed
\ba
\L_{\rm kin} 
& \supset & -\frac{\theta^2 \dot \phi_R^2}{2 \phi_R^2} 
+ \frac{\theta \dot \theta \dot \phi_R}{\phi_R} 
+ \dot h \dot \phi_R
+ \frac12 \dot \phi_R^2 
=
-\frac{\ddot{\phi_R}}{2} \( \frac{\theta^2}{\phi_R}+2 h + \phi_R \) 
\nn \\
&=& \frac12 V_{II}|_{\rm cl} \, (\theta^2 + 2\phi_R h + \phi_R^2).
\label{LkinGB}
\ea
To get the second expression we used partial integration, whereas to
obtain the final result we used the generalized Goldstone theorem
\eref{mI}, which follows from gauge invariance and the background
equations of motion.  The first term in \eref{LkinGB} exactly cancels
the mass term $V_{II}$ in the potential \eref{expV}.  Hence, taking
the system on-shell, all $\theta$-dependent terms can be eliminated,
and in this sense it is still possible to go to unitary gauge.  The
gauge field acquires a mass by eating the massless Goldstone.  The
second term in \eref{LkinGB} cancels the tadpole in the potential.  This
just reflects that even though $\phi_R$ does not minimize the
potential, on-shell it does extremize the action, and thus $\delta
S/\delta \phi_R =0$.  Finally the last term just contributes to the
background energy density, which gets contributions from both kinetic
and potential terms.

Finally we remark that we used the decomposition \eref{flds_exp}  merely for its computational advantages in the next section. To see that the Goldstone boson $\theta$ disappears from the action in unitary gauge, it is easier to use the decomposition $\Phi=\rho \e^{i\theta}$. Here one does not even need to go on-shell to see the Goldstone boson $\theta$ disappear from the action in unitary gauge. 

\subsection{Coleman-Weinberg corrections}

For a theory described by a set of quantum fields of spin $J_i$
Coleman and Weinberg (CW) have calculated the one-loop corrections to
the effective action $\Gamma^{\rm 1-loop} = \int \dd^4 x V_{\rm CW}$, with
\cite{CW}\footnote{Note that the cutoff in this expression is on
  spatial momenta, which explains the difference in coefficients with
  the more common expression with a cutoff on Euclidean 4-momentum.}
\ba
V_{\rm{CW}} &=& \frac12 \sum_{i}  (-1)^{2J_i} (2J_i+1)  \int
\frac{\dd^3k}{(2\pi)^3} \sqrt{k^2 +m_i^2}
\nn\\&=&
\frac{1}{16\pi^2} \sum_{i} (-1)^{2J_i} (2J_i+1) 
\(m_i^2\Lambda^2- \frac14 m_i^4 \ln\(\frac{\Lambda^2}{m^2}\)+\dots\).
\label{CW}
\ea
Here the sum is over all the fields in the model and $\Lambda$ denotes
the energy cut-off. We only kept the relevant (divergent and field-dependent) terms. This expression is valid for a theory in which the
Higgs field is in its minimum, and the background is
time-independent. We now wish to find out how to calculate CW
corrections for an evolving Higgs field.  Based on the discussion in
this section, we would be tempted to use unitary gauge.  If we take
the system on-shell, all reference to the Goldstone boson mass can be
eliminated.  The CW potential is then obtained by summing over the
real part of the Higgs field $h$, and the massive gauge boson. This
procedure, however, leads to problems.

First of all, in a globally supersymmetric theory there are no
quadratic divergences in \eref{CW}, as the bosonic and fermionic
contributions cancel out.  However, here one calculates masses
as second derivatives of the Lagrangian, without demanding the
background Higgs field to be on-shell.  Hence, this calculation also
takes into account a non-zero $m_I^2 = V_{II}|_{\rm cl}$.  If we
remove the Goldstone boson ``by hand'' by going to unitary gauge, this
implies removing the non-zero term in \eref{CW} corresponding to $m_I
\neq 0$.  Consequently, the quadratic divergences would no longer
vanish.  If true, this would have huge consequences for supersymmetric
cosmology. For example, it would be disastrous for supersymmetric
Higgs inflation \cite{einhorn,ferrara1,ferrara2}.

A related problem with removing the Goldstone boson ``by hand'' is
that it gives a discontinuous one-loop potential. When the Higgs field
moves from $\phi_R=0$ to an infinitesimally small amount $\phi_R =
\epsilon$, we go from the symmetric to the broken phase.  The
d.o.f. in the symmetric phase are the real and imaginary parts of the
Higgs $h$ and $\theta$, whereas in the broken phase in unitary gauge
we only have the Higgs $h$ and the massive gauge boson $A_\mu$.
Suddenly the Goldstone boson $\theta$ would not be physical anymore. Its
contribution to the Coleman-Weinberg potential, therefore, should be
omitted, causing a discontinuity in the potential.  This cannot be
correct. 

Therefore we should calculate the Coleman-Weinberg potential \ref{CW}
for a Higgs field displaced from its minimum, in a gauge different
from unitary gauge, and check whether it indeed makes sense to simply
omit the Goldstone boson.

\section{Non-perturbative calculation of the equations of motion} 
\label{berekening}

The previous section's considerations lead us to a careful analysis of
the Coleman-Weinberg corrections to a theory with a displaced Higgs
field.  To take the time-dependence into account we use the
Schwinger-Keldysh or closed-time-path (CTP) formalism
\cite{ctpschwinger,ctpkeldysh,Bakshi1,Bakshi2,ctpjordan,ctpcalzetta,ctppaz,ctpweinberg}. In
this for\-ma\-lism one compares two $\emph{in}$-states rather than an
$\emph{in}$-state and an $\emph{out}$-state. As we are interested in
expectation values at one given point in time, not in transition
amplitudes, it seems more useful to work in this formalism where we do
not need to know the $\emph{out}$-state explicitly. More details on
the CTP formalism can be found in appendix \ref{ctp}.  As it turns
out, the difference between the CTP and the usual S-matrix approach in
the non-perturbative one-loop calculation discussed below vanishes,
and no specific CTP knowledge is needed. This is different for the
perturbative one-loop calculation presented in appendix \ref{pert}.
Our notation and calculation closely follow the work of Heitmann and
Baacke
\cite{baacke1,baacke2,baacke3,baacke4,heitmann,Heitmannmt,Heitmannphdt}. 

Rather then determining the quantum effective action, it turns out
easier to calculate the one-loop corrected equations of motion for the
classical field.  The reason is that the latter can be expressed
directly in terms of the resummed propagator, and as such allows for a
non-perturbative approach.  The equations of motion follow from the
effective action $\Gamma$ in the CTP formalism via $\delta
\Gamma/\delta \phi_+|_{\{J_+ =J_-=0\}} =0$.  Hence, up to a
field-independent constant the effective action can always be obtained
by formally integrating the field equations.

\subsection{Gauge fixing}

To gauge fix the action we use $R_\xi$-gauge.  We add a gauge fixing
term
\be
\L_{\rm GF} = -\frac{1}{2\xi} G^2, \qquad 
G = \partial_\mu A^\mu - \xi g (\phi+h)\theta.
\label{L_GF}
\ee
For notational convenience we dropped the subscript $R$ from the
classical background field. With this choice the term $\propto A^\mu
\partial_\mu \theta (\phi+h)$ in the kinetic terms \eref{expK} is
eliminated.  The corresponding Faddeev-Popov determinant is
\be \L_{\rm FP} = \bar \eta g \frac{\delta G}{\delta \alpha}\eta
= \bar \eta \[- \partial^2 - \xi g^2 (\phi+h)^2 +\xi g^2 \theta^2\] \eta,
\ee
with $\alpha$ the infinitesimal parameter of a $U(1)$ gauge
transformation.  Adding it all together we can write
\be
\L_{\rm tot} = \L + \L_{\rm GF} + \L_{\rm FP} 
=\L_{\rm cl} (\phi) + \L_{\rm free} +  \L_{\rm int}(t).
\label{L_FP}
\ee
The purely classical terms are in $\mathcal{L}_{\rm cl}$. The free
Lagrangian contains the time-independent terms quadratic in the
fluctuation fields, from which the free propagators are constructed.
The interaction Lagrangian contains all other terms, which are treated
as perturbations.  Explicitly,
\ba
\L_{\rm cl} &=& \frac12 \partial_\mu \phi \partial^\mu \phi - V(\phi)
\label{L_cl}
\\ 
 \L_{\rm free}&=& 
-\frac12 A^\mu \[-g_\mn (\partial^2+ g^2 \phi_0^2) 
+ \partial_\mu \partial_\nu(1-\frac1{\xi})\] A^\nu
 -\bar \eta \[ \partial^2 +\xi g^2 \phi_0^2 \]\eta
\nn \\ &&-\frac12 h\[\partial^2+ V_{hh}(0)\]h
-\frac12 \theta\[\partial^2+ V_{\theta\theta}(0)
 +\xi g^2 \phi_0^2\]\theta
\label{L_free} \\
\L_{\rm int} &=& -  h \[ \partial^2\phi + V_{\phi}\]  
+\frac{g^2}{2} (\phi^2 - \phi_0^2)
\[A_\mu A^\mu- \xi \theta^2 -2 \xi \bar \eta \eta \] 
\nn \\ && -2g \partial_\mu \phi A^\mu \theta - \frac12 (V_{hh}(t) - V_{hh}(0))h^2
 - \frac12 (V_{\theta\theta}(t) - V_{\theta\theta}(0))\theta^2
+...,
\label{L_int}
\ea
with $\phi_0 = \phi(0)$ the initial field value.  The ellipses denote
terms of third or higher order in the fluctuation fields.  As before
$V_\phi = \partial_\phi V$ etc. with $V$ the classical potential.

We define the ``mass''-matrix via
\be m^2_{\alpha\beta} = - \frac{\partial^2 \L}{\partial \chi_\alpha
\partial \chi_\beta} = \bar m_{\alpha\beta}^2 + \delta m^2_{\alpha \beta}(t)
, \qquad \chi_\alpha = \{A^\mu,\eta,h,\theta\}
\ee
which can be split in a free time-independent part, denoted by an
overbar, and a time-dependent part.  The non-zero elements of the
mass matrix are:
\ba
&& m_{A^\mu A^\nu}^2 = -g^2 \phi^2 g^\mn \equiv -m_A^2 g^\mn, \quad
m_\eta^2 = \xi g^2 \phi^2, \quad
m_h^2 = V_{hh}, \quad
m_\theta^2 = V_{\theta\theta} + \xi g^2 \phi^2, \nn \\
&&m^2_{\theta A^\mu} = 2g \dot \phi \delta_0^\mu 
\equiv m^2_{A\theta}\delta_0^\mu, 
\label{mass}
\ea
where for the diagonal entries we used the notation $m^2_\alpha =
m^2_{\alpha \beta} \delta^\alpha_\beta$.  The only off-diagonal term
is the term in the second line above mixing the Goldstone boson and the
temporal part of the gauge field.  The temporal gauge boson has a
wrong sign mass. As it also has a wrong sign kinetic term, the
dispersion relation for $A^0$ is still of the standard form
$\omega^2_{A^0}= \vec k^2 + |m_{A^0A^0}^2|= \vec k^2 + m_{A}^2$.

From the interaction Lagrangian we can read off the n-point functions.
Of particular interest for the calculations performed in this paper
are the following one-point, two-point and three-point vertices:
\be
\Gamma_h = -i(\Box \phi + V_\phi), \qquad
\Gamma_{\alpha\beta} = -i m^2_{\alpha \beta}, \qquad
\Gamma_{h\alpha \beta} = 
-i\partial_\phi
m^2_{\alpha \beta}.
\label{vertex}
\ee 
%

\subsection{Real scalar field} \label{rsf}

\begin{figure}
\begin{center}
\includegraphics[scale=1]{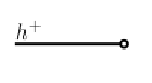}
\caption{The classical background field equation can be derived from
  the tadpole diagram with one external $h^+$ leg as shown in the
  figure.  The cross represents the one-point function.}
\label{F:tadpole}
\vspace{1cm}
\includegraphics[scale=1]{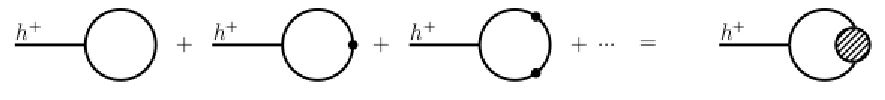}
\caption{The background field equation of motion is corrected by all
  one-loop diagrams with one external $h^+$ leg as shown in the
  figure.  The lines correspond to the bare propagator, and the dots
  to two-point (mass) insertions.  This can be resummed to get a resummed
  propagator running in the loop, depicted here by a blob.}
\label{F:loop}
\end{center}
\end{figure}

To warm up, we first perform the one-loop calculation for a single
real scalar field rolling down the potential, using the
Schwinger-Keldysh formalism. We wish to calculate the one-loop
correction to the equation of motion, defined as the sum over all
one-loop diagrams with one external leg (of the quantum field $h$) shown in Figure \eref{F:loop}.
Integrating the equations of motion, we retrieve the standard
Coleman-Weinberg potential \eref{CW} in the time-independent limit
$\dot \phi(t) =0$.  We follow the treatment in \cite{baacke1,baacke2}.

Consider a real scalar field expanded around a classical field value
$\Phi = \phi(t) + h(x,t)$, where we can split $\phi(t) = \phi_0 +
\delta \phi(t)$ with $\delta \phi(0) =0$. The one-loop correction to
the equation of motion comes from the sum of all vacuum loops with one
external leg, as depicted in Figure \eref{F:loop}, which can be expressed as
\cite{ctpcalzetta,ctppaz,Heitmannmt,Heitmannphdt}
\ba
0&=&\frac{\delta \Gamma}{\delta \phi_+}\bigg|_{\{J_+ =J_-=0\}} =
i\(\Gamma_h +   \frac12\Gamma^+_{hhh} G^{++}_h(0)\) \bigg|_{\{J_+ =J_-=0\}}
+\O(h^2)\nn\\
&=&\Box \phi + V_\phi +  \frac12 (\partial_{\phi}m^2_{hh})
G_h^{++}(0)
+\O(h^2),
\label{eom_scalar}
\ea
where in the second line we have set $\phi_+|_{\{J_\pm = 0\} }= \phi$.
$G_h^{++}(x,x')$ is the dressed or resummed propagator taking all
two-point insertions into account, which is defined in appendix
\ref{dressed}.  The first term is the classical tree level
contribution to the equation of motion, which can be found from the
tadpole diagram in Figure \eref{F:tadpole}.  The second term is the
1-loop correction, which is the sum of all one-loop diagrams with one
external $h^+$ leg and arbitrary number of mass insertions, as
depicted in Figure \eref{F:loop}.  The factor $1/2$ is a symmetry
factor originating from the reflection symmetry of the Feynman
diagrams.  All the relevant n-point functions are defined in
\eref{vertex}.

We only need to consider the 1-loop contribution on the $(+)$-branch
of the Schwinger-Keldysh in-in formalism (the calculation on the
$(-)$-branch gives the same result).  Therefore the calculation is
fully analogous to the usual in-out scattering matrix calculation.
For ease of notation we drop the $(++)$ superscript in the following.

The dressed propagator can be expressed in terms of the mode functions
\be
G_h(0) = \int \frac{\dd^3k}{(2\pi)^3} \frac{|U_{h}|^2}{2 \bar \omega_{h}},
\label{GU}
\ee
where for notational convenience we dropped the subscript $\vec k$ on
the mode functions and the frequency.  The mode functions satisfy a
wave equation with a time-dependent frequency (which can be read off
from the quadratic part of the Lagrangian --- see appendix \ref{ctp}
for more details)
\be
\[ \partial_t^2 + \omega_{\vec k,h}^2\] U_{h}(t) =0,
\qquad {\rm with} \; \; U_{h}(0) = 1, \;\; 
\dot U_{h}(0) = - i \bar \omega_{h}.
\label{mode_h}
\ee
The frequency can be split in a time-independent and a time-dependent
piece $\omega_{h}^2(t) = \bar \omega_{h}^2 + \delta m_h^2(t)$ with
$\bar \omega_{h}^2 = \vec k^2 + \bar m_h^2$, with as before the
overbar denoting the time-independent quantities.  To solve the mode
equation \eref{mode_h} we make the Ansatz
\be
U_{h} = \e^{-i \bar\omega_{h} t}(1+ f_h(t)).
\ee
The function $f_h$ satisfies $\ddot f_h-2i\bar\omega_h \dot f_h=
-\delta m_h^2(1+f_h)$ and has boundary conditions
$\dot{f_h}(0)=f_h(0)=0$.  This can be solved using the Green's
function method to yield:
\be f_h = -\frac{1}{\bar\omega_h} \int_0^t \dd t' \sin(\bar\omega_h \Delta
t)\e^{i\bar \omega_h \Delta t} (1 + f_h(t')) \delta m_h^2(t'),
\ee
with $\Delta t = t -t'$. We can solve the mode equations iteratively
order by order in mass insertions $f = f^{(1)} + f^{(2)} +...$.  To
isolate the divergent part it is enough to only go to first order,
since $|U_h|^2 = 1 + 2 {\rm Re} f_h^{(1)} + \O(k^{-4})$ --- since, as
we will see in a moment, for large momentum $f_h^{(1)} \propto
k^{-2}$. For $f_h(t) = 0$ we get back the bare (free) propagator with
no mass insertion.\footnote{Note that $f_h(t) = 0$ corresponds to the
first order result in the perturbative calculation of appendix
\ref{pert}, while $f_h^{(1)}$ corresponds to the second order result
in the perturbative calculation.}  Define $f_h^{(1)}$ as the first order correction
in the mass insertion; it is given by
\be f_h^{(1)} = -\frac{1}{\bar\omega_h} \int_0^t \dd t' \sin(\bar\omega_h \Delta
t)\e^{i\bar\omega_h \Delta t} \delta m_h^2(t').
\ee
Using partial integration, and taking the real part gives
\be
{\rm Re}  f_h^{(1)} 
= - \frac{\delta m_h^2(t)}{4\bar\omega_h^2} + 
\frac{1}{4\bar\omega_h^2} \int \dd t' \cos(2\bar\omega_h \Delta t)
\partial_{t'}(\delta m_h^2(t'))
= - \frac{\delta m_h^2(t)}{4\bar\omega_h^2} + \O(\bar\omega_h^{-3}).
\ee
Finally
\ba  G_h(0)
&=&  \int \frac{\dd^3k}{(2\pi)^3}
\frac{1+ 2 {\rm Re}f_h^{(1)}+...}{2\bar \omega_h} = 
\frac{1}{4\pi^2} \int k^2
\dd k \(\frac{1}{k} - \frac1{2k^3}(\bar m_h^2+ \delta m_h^2) + \O(k^{-5}) \)
\nn\\ &=& \frac{1}{8\pi^2} \( \Lambda^2 - \frac12 m_h^2(t) \ln \(\frac{\Lambda^2}{m_h^2}\) \)
+{\rm finite}.
\label{Vh}
\ea
The 1-loop equation of motion \eref{eom_scalar} thus becomes
\be
0=\Box \phi + V_\phi + \frac{\partial_\phi m^2_{h}}{16\pi^2} \( \Lambda^2 
- \frac12 m_h^2(t) \ln \(\frac{\Lambda^2}{m_h^2}\) \).
\ee
We can integrate the last term with respect to $\phi$ to get the
one-loop correction to the effective potential.  Up to field-independent
and finite terms:
\be
\Gamma^{1-{\rm loop}}= -\int \dd^4 x \int \dd \phi 
\frac{\partial_{\phi}m^2_h}{16\pi^2} 
\( \Lambda^2 - \frac12 m_h^2 \ln \(\frac{\Lambda^2}{m_h^2}\) \) 
= -\frac{1}{16\pi^2} \int \dd^4 x\(m_h^2 \Lambda^2
 -\frac14 m_h^4 \ln\(\frac{\Lambda^2}{m_h^2}\)
\)  .
\ee
In the static limit that the background field, and thus the mass term,
is time-independent $\Gamma^{1-{\rm loop}} = -\int \dd^4x V_{CW}$, and
we recover the Coleman-Weinberg result \eref{CW}.


\subsection{Abelian Higgs model}

We now extend the analysis to a U(1) model with a complex Higgs field.
The one-loop equation of motion \eref{eom_scalar} generalizes to
\be
0=\Box \phi + V_\phi +  \frac12 \left(\partial_\phi m^2_{\alpha \beta} \right)
G_{\alpha \beta}^{++}(0).
\label{eom}
\ee
We use the Feynman-'t Hooft gauge $\xi = 1$, for which the equations
of motion of $A^i$ and $A^0$ decouple. All four components of the
gauge field satisfy a Klein-Gordon equation.  The quadratic terms for
$\alpha =\{h,\eta,A^i\}$ are diagonal, and the one-loop calculation
proceeds a\-na\-lo\-gous\-ly to the scalar field case discussed in the previous
subsection.  On the other hand, the fields $\{A^0, \theta \}$ couple
in the quadratic terms, because of the non-diagonal mass term
$m^2_{A^0 \theta} \neq 0$, and need to be treated with care.  

The calculation of the propagator for the real scalar $h$ was done in
the previous subsection.  Also for $\eta$, which is an anti-commuting
complex scalar, the scalar field result applies with a factor 2 for
the 2 real d.o.f. and a minus sign to take into account the
anti-commuting nature.   In the $\xi =1$ gauge the propagator $G_{A^i}$ satisfies
$[\Box +m_A^2]G_{A^i} = -i\delta(x-x')$. As this equation is of the
same form as the one for the scalar field propagator, the scalar field
results can be applied.  Hence, $A^i$ contributes as three scalars
with mass $m_A$ each. The result thus is
\ba 
\sum_{\{h,\eta,A^i\}} \frac12 (\partial_\phi m_\alpha^2) G^{++}_\alpha(0)
= 
\frac{1}{16\pi^2} \bigg[&& \left(\partial_\phi m_h^2\right)
\(\Lambda^2 - \frac12 m_h^2
\ln\(\frac{\Lambda^2}{m_h^2}\) \) \nn \\
-{2} &&\left(\partial_\phi m_\eta^2\right) \(\Lambda^2 - \frac12 m_\eta^2
\ln\(\frac{\Lambda^2}{m_\eta^2}\) \)
\nn \\ +{3} &&\left(\partial_\phi m_{A^i}^2\right) \(\Lambda^2 - \frac12 m_{A^i}^2
\ln\(\frac{\Lambda^2}{m_{A^i}^2}\) \)
\bigg].
\ea

The difficulty is in calculating the propagators for $\{A^0,\theta\}$,
as these fields couple in their equations of motion.  We only outline
the calculation, more details can be found in
\cite{baacke1,baacke2}. We define two sets $\alpha = \{1,2\}$ of mode
functions which satisfy (following from the quadratic part of the
Lagrangian, see appendix \ref{ctp})
\be
\biggl[ \biggl( \begin{array}{ccc} 
-\left(\partial_t^2+\bar{\omega}_A^2\right) & 0 \\ 0 & 
\partial_t^2+\bar{\omega}_\theta^2 \end{array} \biggr) +
\biggl( \begin{array}{ccc} -\delta m_A^2 & \delta m_{A\theta}^2 \\ 
\delta m_{A\theta}^2 & \delta m_\theta^2 \end{array}\biggr) \biggr] 
\biggl( \begin{array}{ccc} U_A^\alpha \\ U_\theta^\alpha \end{array} \biggr) =0,
\ee
with
\be
U^\alpha_m (0) = \delta^\alpha_m, \qquad 
\dot U^\alpha_m(0) = -i \bar\omega_m \delta^\alpha_m.
\ee
$ \delta m_m^2$ and $ \delta m_{mn}^2 $ correspond to the diagonal and
off-diagonal entries of the time-dependent part of the mass
matrix. For example: $\bar{m}_A^2=g^2 \phi_0^2$, $\delta
m_A^2=g^2\left(\phi^2-\phi_0^2\right)$. The frequency for the temporal
gauge field is $\omega_{A}^2 = k^2 + m_A^2$.  The $\alpha =1$ mode is
the ``mostly gauge boson'' mode, and $\alpha =2$ is the ``mostly
Goldstone boson mode''.  The modes do not decouple because of the
off-diagonal $\delta m_{mn}^2$ term.  The resummed equal-time
propagator in terms of the mode functions is
\be
G_{kn}(0) = \int\frac{d^3k}{(2\pi)^3}\left[
-\frac1{4\bar\omega_A} \( U^1_k U^{1*}_n + U^{1*}_k U^1_n \)
+ \frac1{4\bar\omega_\theta} \( U^2_k U^{2*}_n + U^{2*}_k U^2_n \)\right]
\ee
and thus
\ba
\sum_{\{\theta,A^0\}} \frac12 \left(\partial_\phi m^2_{\alpha\beta}\right) G^{++}_{\alpha \beta}
=
\frac12\! &&\int \frac{\dd^3k}{(2\pi)^3}  \bigg[ \partial_\phi m_A^2 
\( \frac1{2\bar\omega_A} |U^1_A|^2 \! -\!  \frac1{2\bar\omega_\theta} |U^2_A|^2\)
\nn\\
+&&\partial_\phi m_\theta^2 
\( \frac1{2\bar\omega_\theta} |U^2_\theta|^2 \! -\!  \frac1{2\bar \omega_A} |U^1_\theta|^2\)
\nn \\ +&&
 2 \partial_\phi m^2_{\theta A} 
 \(
 -\frac1{4\bar\omega_A} (U^1_A U^{1*}_\theta+ U^{1*}_A U^1_\theta)
+  \frac1{4\bar\omega_\theta} (U^2_A U^{2*}_\theta+ U^{2*}_A U^2_\theta\)
\bigg]. \nn\\
\label{Gmix}
\ea
To solve for the mode functions make the Ansatz which is consistent
with the boundary conditions if we again choose $f(0)=\dot{f}(0) =0$:
\ba
&&U^1_A = \e^{-i \bar\omega_A t} (1+f_A^1), \qquad
U^1_\theta  = \e^{-i \bar\omega_\theta t} f_\theta^1, \nn \\
&&U^2_\theta  = \e^{-i \bar\omega_\theta t} (1+f_\theta^2), \qquad
U^2_A = \e^{-i \bar\omega_A t} f_A^2.
\label{VmixU}
\ea
We can again solve iteratively, and define an expansion in terms of
mass-term insertions $ f^\alpha_m = f^{\alpha(1)}_m + f^{\alpha(2)}_m
...$.  To isolate the divergent part of the one-loop potential we
again only need the first order result.  Plugging the Ansatz
\eref{VmixU} in the mode equations gives
\ba
&&\ddot f^{m(1)}_\alpha -2i \bar\omega_\alpha \dot f^{m(1)}_\alpha 
= - \delta m^2_\alpha, \hspace{3.7cm}
{\rm for} \;\{m,\alpha\} = \{1,A\},\,\{2,\theta\}\nn\\
&&
\ddot f^{m(1)}_\alpha -2i \bar\omega_\alpha\dot f^{m(1)}_\alpha =
(-1)^m \delta m^2_{A\theta} \e^{(-1)^m i(\bar\omega_A-\bar\omega_\theta)t}
,\quad
{\rm for} \;\{m,\alpha\} = \{1,\theta\},\,\{2,A\}
\ea
where we only kept the highest order results. To do so we used that at
large momentum $\omega^n \partial^m_t f^{(l)} \propto k^{m+n-2l}$.
Just as in the scalar field case, the equations can be solved using the
Green's function method.  The $f^1_A$ and $f^2_\theta$ equations are
exactly the same as found for the scalar in the previous subsection, and
hence give the same result: 
\ba
f^{m(1)}_\alpha &=& - \frac{1}{\bar \omega_\alpha}\int^t \dd t' 
\sin(\bar\omega_\alpha \Delta t)
\e^{i \bar \omega_\alpha \Delta t} \delta m_\alpha^2(t'),
\hspace{3.7cm}
{\rm for} \;\{m,\alpha\} = \{1,A\},\,\{2,\theta\},
\nn \\
f^{m(1)}_\alpha &=&  \frac{(-1)^m}{\bar \omega_\alpha}\int^t \dd t' 
\sin(\bar\omega_\alpha \Delta t)
\e^{i \bar\omega_\alpha \Delta t} \e^{ (-1)^m i(\bar \omega_A -\bar \omega_\theta)t'}
\delta m_{A\theta}^2(t'),\qquad
{\rm for} \;\{m,\alpha\} = \{1,\theta\},\,\{2,A\}. \nn\\
\ea

Now consider the first line of \eref{Gmix}.  The terms $|U_A^2|^2 =
|f^{2(1)}_A|^2$ and $|U_\theta^1|^2 = |f^{1(1)}_\theta|^2$ are second
order in $f$ and thus give no contribution to the divergent terms.
The remaining terms on this line are analogous to the scalar loop,
they correspond to Feynman diagrams with $\theta$ and $A^0$ loop
running in the loop, and give the standard Coleman-Weinberg result.
Hence we get a contribution as in \eref{Vh} but now for $\theta,A^0$.
Remains to evaluate the second line of \eref{Gmix}:
\ba
&&{\partial_\phi m^2_{A\theta}} \int \frac{\dd^3k}{(2\pi)^3} \(
 -\frac1{4\bar\omega_A} 2 {\rm Re}[ \e^{it(\bar\omega_A -\bar\omega_\theta)} 
f_\theta^{1(1)}] + ( \{1,A\} \leftrightarrow \{2,\theta\}) \)\nn \\
&=& \frac{\partial_\phi m^2_{A\theta}}{4} \int \frac{d^3k}{(2\pi)^3}  
\biggl[ \frac{\cos{[(\bar\omega_A-\bar\omega_\theta)t]}}{\bar\omega_A\bar\omega_\theta} 
\int_0^{t'} dt' 
\sin{[2\bar\omega_\theta\Delta t]}  
 \cos{[(\bar\omega_\theta-\bar\omega_A)t']}\delta m^2_{A\theta}(t')
+(a \leftrightarrow \theta) \biggr]
 \nn\\
&=&\frac{\partial_\phi m^2_{A\theta}}{8} \int \frac{\dd^3k}{(2\pi)^3} 
 \frac1{\bar\omega_A\bar\omega_\theta(\bar\omega_A+\bar\omega_\theta)}
\cos^2{[(\bar\omega_A-\bar\omega_\theta)t]} \delta m_{A\theta}^2(t)  
+\O(\omega^{-4})\nn \\
&=&  
\frac{\partial_\phi m_{A\theta}^2}{4\pi^2}  m_{A\theta}^2
\ln \(\frac{\Lambda^2} {m_{A\theta}^2}\) +{\rm finite}.
\label{Vmix}
\ea
To obtain the third line we used partial integration. Further
$m_{A\theta}^2= \delta m_{A\theta}^2$, as there is no time-independent
mix term.

Adding it all up, the one-loop equation of motion thus becomes
\be
0=\Box \phi + V_\phi +\frac{\partial_\phi m_{A\theta}^2}{4\pi^2}  m_{A\theta}^2
\ln \(\frac{\Lambda^2} {m_{A\theta}^2}\) +
\sum_{\{\phi,\eta,A^i,\theta, A_0\}} 
\frac{S_i}{16\pi^2}  \partial_\phi m_i^2
\(\Lambda^2 - \frac12 m_i^2 \ln\(\frac{\Lambda^2}{m_i^2}\) \)
\label{eom2}
\ee
with $S_i =\{1,-2,3,1,1\}$ for $i=\{\phi,\eta,A^i,\theta,A_0\}$
counting the degrees of freedom.  Since the equations of motion follow
from the functional derivative of the effective action, one can invert
the process, and find (up to a field-independent constant) the
effective action by integrating the field equations with respect to
$\phi$:
\ba
\Gamma^{1-{\rm loop}} &=& 
-\frac{1}{16\pi^2} \int\dd^4 x \bigg[ \Lambda^2 \(
m_h^2 - 2m_\eta^2+3 m_{A^i}^2 + m_\theta^2 + m_{A_0}^2\) 
\label{result_np}
 \\ &&- \frac 14 \ln \Lambda^2 \(
m_h^4 - 2m_\eta^4+3 m_{A^i}^4 + m_\theta^4 + m_{A_0}^4
-2 m_{\theta A}^4 \)\bigg]+{\rm finite}
\nn \\ &=&
-\frac{1}{16\pi^2} \int\dd^4 x\bigg[ \Lambda^2 \(V_{hh}+V_{\theta\theta}+3 m_A^2\)
-  \frac14\ln \Lambda^2 \(V_{hh}^2+V_{\theta\theta}^2+3 m_A^4
-6  V_{\theta \theta} m_A^2 \) \bigg].\nn 
\ea
In the second step we used the zeroth order background equation of motion and gauge
invariance to write 
\be
\int \dd t \, m_{\theta A}^4= 4g^2 \int \dd t \dot \phi^2 
= - 4g^2 \int \dd t \phi \ddot \phi = 4g^2 \int \dd t \phi V_\phi 
= 4 \int \dd t \,m_A^2 V_{\theta \theta}
\label{eom_zero}
\ee
up to higher loop corrections.  With this substitution the final
expression is in terms of explicitly gauge independent quantities.
This can be seen more explicitly in the perturbative calculation in
appendix A, which is done for arbitrary gauge parameter $\xi$.  As a
result the {\it on-shell} one loop effective potential is gauge
invariant.  In the static limit $V_{\theta \theta} \to 0$ and all
other masses are time-independent, our results reproduce the standard
Coleman-Weinberg potential \eref{CW}.

The gauge independent part of the Goldstone boson mass $V_{\theta
  \theta}$ appears explicitly in the one-loop potential.  Except for
the very last term in \eref{result_np}, the one loop potential can be
obtained from the Coleman-Weinberg potential, treating $\theta$ as a
physical bosonic degree of freedom.  The calculation done in unitary
gauge with $\theta$ completely ``gauged away'' from the potential
(which, as discussed in subsection \ref{hm}, for $\phi$ displaced away
from its minimum seems only possible on-shell) gives the wrong answer.
This answers the question posed at the beginning of this section.  The
Goldstone boson cannot be removed ``by hand'', and keeping its
contribution in the one-loop potential assures this is continuous.

Our answers disagree with the naive expectation obtained in unitary
gauge, where the Goldstone boson is absent. The reason is that unitary
gauge is a singular limit. It corresponds to taking the limit $\xi \to
\infty$ such that the $\theta$ propagator vanishes. This procedure,
however, does not commute with the $k \to \infty$ limit taken in the
momentum integrals to isolate the divergent terms. That unitary gauge
gives an incorrect result has been noted before \cite{jackiw}.  In
this gauge higher order loop corrections affect the leading term and
must be taken into account \cite{braaten}.

The last term on the last line of \eref{result_np} can be interpreted as
a correction to the Coleman-Weinberg potential, due the fact that
$\phi$ is rolling down its potential rather than sitting in its
minimum.  It vanishes in the static limit; note in this respect that
it came from the $\dot \phi$ term.


\subsection{Fermions}

Even if the focus in this article is obviously on scalar fields, we
want to include a section on fermionic fields here, in order to arrive
at a more complete picture of one-loop corrections in a theory
with a displaced Higgs field. In Standard Model Higgs inflation the
top quark contributes significantly to the one-loop potential,
whereas in supersymmetric theories Higgsinos and gauginos should be
taken into account as well.  The full calculation for fermions has
been done in \cite{baacke3}. Here we summarize their results, adapted to
calculate the effective potential.

In a supersymmetric theory, the gauginos and Higgsinos couple in the
mass matrix if the gauge symmetry is broken.  It is always possible to
diagonalize the mass matrix, and do the calculation in terms of mass
eigenstates, whether the theory is supersymmetric or not. There are no
mixed loops, such as in the bosonic sector, where the Goldstone boson
and temporal gauge field are coupled. In the static limit, the
one-loop is given by the Coleman-Weinberg potential \eref{CW}, to
which each mass eigenstate contributes. To find possible
time-dependent corrections, one can again use the CTP formalism. 
We calculate the one-loop correction to the equation of motion for the
background field due to a fermion loop.  Only fermions which have a
field dependent mass term contribute.

Consider a Dirac or Majorana fermion with Lagrangian
\be
\L = \bar \psi (i \gamma^\mu \partial_\mu - m_\psi(t)) \psi.
\ee
For a Yukawa type interaction the fermion mass is $m_F = \lambda
\phi$, with $\lambda$ the Yukawa coupling and $\phi$ the Higgs field.
The one-loop equation of motion for the background field is \cite{baacke3}
\ba
0&=&\Box \phi + V' 
+  \frac12 \sum_{\rm bosons}  \left(\partial_\phi m_{\alpha\beta}\right) G_{\alpha\beta}^{++}(x,x)
-  \frac12 \sum_{\rm fermions} \left(\partial_\phi m_{\alpha\beta}\right) G_{\alpha\beta}^{++}(x,x).
\label{eom_h}
\ea
The bosonic contribution was calculated in the previous section, here
we concentrate on the second fermionic contribution.  For simplicity
we do the calculation for a single fermion field. The equal-time
dressed propagator for a fermion is given in the appendix
\eref{G_psi}.  The Dirac equation can be rewritten as a second order wave
equation, using a particular Ansatz for the spinors
(\ref{Ansatz},~\ref{mode_psi}).  This maps the problem to an
equivalent form as for the real scalar discussed in section \ref{rsf}.
The one-loop potential can be calculated analogously.  The result
found in \cite{baacke3} is
\be
-\sum_{\rm d.o.f.} \frac12 \left(\partial_\phi m_\psi\right) G_\psi^{++}(0)= -
\sum_{\rm d.o.f.}\frac{\partial_\phi m_\psi}{8 \pi^2}
\[ m_\psi \Lambda^2 -\frac12 \( m_\psi^3 + \frac12\ddot m_\psi\) \ln \(\frac{\Lambda^2}{m_\psi^2}\)\],
\ee
where the sum is over all helicity states, 4 for a Dirac fermion and 2
for a Majorana/Weyl fermion. For a Yukawa mass we have on-shell
\be
\frac{\ddot m_\psi}{m_\psi} = \frac{\ddot \phi}{\phi} = -V_{\theta\theta}.
\ee
Integrating the field equations to get the the one-loop correction to
the effective action gives
\be
\Gamma^{1-{\rm loop}} = 
\sum_{\rm d.o.f.} \frac{1}{16\pi^2} \int\dd^4 x \bigg[ m_\psi^2\Lambda^2 
-\frac{1}{4}(m_\psi^4-m_\psi^2 V_{\theta\theta}) \ln \Lambda^2\bigg] + {\rm finite}.
\ee
In the static limit $V_{\theta\theta} \to 0$, this indeed reproduces
the standard CW result \eref{CW}.  We thus find that the
time-dependent corrections scale with the Goldstone boson mass.


\section{Conclusions and outlook} \label{conclusions}

In this work we have computed the one-loop corrected equations of
motion for the background Higgs field, and, by integrating, the one-loop
effective action  for a theory in which the Higgs field is slowly rolling down its
potential. For our U(1) toy model with a complex Higgs field $\Phi=\phi_0
+\delta\phi(t)+h(x,t) + i\theta(x,t)$ moving through a potential $V$
and a vector field $A^\mu(x,t)$ we find 
\ba
\Gamma&=&\int \dd^4 x 
\left\{ \L_{\rm cl} -
\frac{1}{16\pi^2} \bigg[ \Lambda^2 \(V_{hh}+V_{\theta\theta}+3 m_A^2\)
-  \frac14\ln \Lambda^2 \(V_{hh}^2+V_{\theta\theta}^2+3 m_A^4
-6  V_{\theta \theta} m_A^2 \) \bigg] \right\}
\nn \\ &+&  \O(\hbar^2).
\label{result2}
\ea
up to finite and field-independent terms.  To write the results in
this manifestly gauge invariant way we used the zeroth order
background equations of motion and gauge invariance to replace
$m_{A\theta}^4 \to m_A^2 V_{\theta \theta}$ in the one-loop
correction. The potential is completely arbitrary.  We first remark
that in the static case one has $V_{\theta \theta}=0$ and we are left
with the well-known Coleman-Weinberg result.  Note that the last term
in \eref{result2} can change the sign of the log term, but only if all
masses are of the same order.  If the scalar and gauge boson masses
are hierarchical, it will be negligible.  This may be important for
Higgs inflation in certain GUT models.

With the Higgs field displaced from its minimum, the Goldstone boson
$\theta$ is massive. It cannot be removed from the theory. At the
classical level we can still use unitary gauge (and the equation of
motion) to eliminate the Goldstone boson from the theory, at the
quantum level this procedure gives wrong results.  In particular, the
Goldstone boson still contributes to the effective action as if it was
a massive scalar degree of freedom.  This comes in addition to the
contribution from the massive gauge boson.  Thus even if we should not
call the Goldstone boson ``physical'' (its associated degree of
freedom, after all, has been used to give the gauge boson a mass), the
factors of $V_{\theta\theta}$ in the potential are real and can not be
discarded.  (One might argue that they are induced by the massive
gauge boson.)

The equivalent calculation performed in unitary gauge gives wrong
answers.  The reason is that unitary gauge is ill-defined.  It
corresponds to taking the limit $\xi \to \infty$ such that the
$\theta$ propagator vanishes. This procedure, however, does not
commute with the $k \to \infty$ limit taken in the momentum integrals
to isolate the divergent terms. Problems with unitary gauge were
noted before, for example in the calculation of the one-loop potential
at finite temperature \cite{jackiw}. In that context it was shown that
two-loop effects contribute at the same order, and cannot be neglected
\cite{braaten}.

Our results imply that supersymmetric Higgs inflation is free of
quadratic divergencies, as the bosonic and fermionic degrees of
freedom still cancel.  In addition the effective potential is
continuous in going from the symmetric to the broken phase, as it
should be.

Our calculations closely followed the work of Heitmann and Baacke,
generalized to an arbitrary potential.  Moreover we explicitly show
that the results are gauge invariant on shell. Our results reproduce
the Coleman-Weinberg results in the static limit.
Ref. \cite{boyanovski} has calculated the effective potential in terms
of manifestly gauge invariant quantities, but only in the adiabatic
limit, which does not take into account the time-dependence of the
rolling Higgs field. These time-dependent corrections are essential
for us to show the gauge-independence of the final result.

To get from our toy model to the case of Higgs inflation the first
step is to generalize the gauge group U(1) to the Standard Model or
GUT gauge group, depending on the inflation model under consideration.
This is a trivial extension of our results. The second, far less
trivial, step is to do the calculation in a Friedmann-Robertson-Walker
spacetime rather than in Minkowski spacetime. The scalar and fermion
field contributions can rather straightforwardly be generalized, and
yield additional corrections to the Coleman-Weinberg potential due to
the expansion of the universe.  But the difficulties arise in the
gauge boson and Goldstone boson sector.  In a cosmological spacetime
Lorentz symmetry is broken, and as a consequence the temporal and
longitudinal/transversal parts of the gauge field no longer decouple.
This is left for future work.

A third step left to be done is generalizing the results to
non-canonical kinetic terms. If the kinetic terms cannot be
diagonalized by simple field redefinitions, as is the case in Standard
Model Higgs inflation, the radial Higgs field and Goldstone bosons
couple in a non-trivial way.  The equations can still be solved in the
adiabatic approximation. However, different approximation schemes have
to be developed if the field evolution is fast, which is the case
after inflation.

\section*{Acknowledgments}
The authors are supported by a VIDI grant from the Dutch Science
Organization FOM.  We thank Mikhail Shaposhnikov, Damien George,
Jan-Willem van Holten, Eric Laenen and Jan Smit for useful
discussions. We are very grateful to Katrin Heitmann for sending us
her master's thesis.

\appendix


\section{CTP formalism} \label{ctp}

In the usual S-matrix approach, also called in-out formalism, the
generating functional describes the transition from an in-state vacuum
in the past to an out-state vacuum in the future $Z[J]=\langle
0,t_{\rm in}|0, t_{\rm out} \rangle_J$, which is calculated in the
presence of an external source $J$.  In the path-integral formulation
\be
Z[J] = \int \D \phi \, \e^{i S[\phi] + \int \dd^4 x J \phi}.
\ee
This formalism is well suited to calculate scattering amplitudes,
processes in which the out-state is known.  In non-equilibrium
situations it is more useful to calculate the physically relevant
field expectation values of an observable $\langle 0,t_{\rm in} |
\O|0, t_{\rm in} \rangle$ taken with respect to the same states. The
generating functional in this in-in formalism, also known as
Schwinger-Keldysh or closed time-path (CTP) formalism
\cite{ctpschwinger,ctpkeldysh,Bakshi1,Bakshi2,ctpjordan,ctpcalzetta,ctppaz,ctpweinberg}, is defined employing two external sources:
\be
Z[J^+,J^-] = _{J_-} \langle 0,t_{\rm in}|0,t_{\rm in} \rangle_{J_+}
= \sum_\alpha  \langle 0,t_{\rm in}|\alpha,t_{\rm out} \rangle_{J_-}
 \langle \alpha,t_{\rm out}|0,t_{\rm in} \rangle_{J_+},
\ee
where the sum goes over a complete set of out states. The above
expression can be understood as the in-vacuum going forward in time
under influence of the $J_+$ source, and then returning back in time
under the influence of the $J_-$ source.  On both branches propagators
and vertices can be defined, with the $-$-branch giving the time
reversed of expressions the $+$-branch.

\subsection{Free propagators}

We will define free propagators and vertices, needed for the one-loop
perturbative calculation.  The free Lagrangian \eref{L_free} is of the
form $\L^{\rm free}=-(1/2)\sum_i \chi_i(x^\mu) \bar K^i (x^\mu)
\chi_i(x^\mu)$, with the sum over all (bosonic) fields $\chi_i =
\{h,\theta,\eta,A^\mu\}$. The time-dependent parts of the quadratic
action are treated as interactions.  As before, the overbar denotes
that we only consider the time-independent parts of the quadratic
terms. The free propagators are defined as
\be
\( \begin{array}{cc}
\bar K^{i}(x^\mu) & 0 \\
0 & -\bar K^{i}(x^\mu)
\end{array} \)
\( \begin{array}{cc}
\bar G_{i}^{++}(x^\mu - y^\mu) & \;\bar G_{i}^{+-}(x^\mu - y^\mu) \\
\bar G_{i}^{-+}(x^\mu - y^\mu) & \;\bar G_{i}^{--}(x^\mu - y^\mu)
\end{array} \)
= -i \delta(x^\mu - y^\mu) \mathbf{I}_{2}.
\label{defG}
\ee
These equations can be easily solved in Fourier space, for example the
$(++)$ Green's function is
\ba
\bar G_{i}^{++}(k) &=& \frac{i}{k^2-\bar m_i^2+i\eps} \nn\\
(\bar G_{A}^{++})_\mn(k) &=& -\frac{i}{k^2-\bar m_A^2+i\eps} 
\(g_\mn - \frac{k_\mu k_\nu}{k^2}\)
-\frac{i\xi}{k^2-\bar m_\xi^2+i\eps} \(\frac{k_\mu k_\nu}{k^2}\),
\ea
where the first expression applies to the scalars $i
=\{h,\theta,\eta\}$, and the second to the vector boson. 
 Here the masses correspond to the {\it time-independent}
parts of the mass terms \eref{mass}, indicated by the overbar,
appearing in $\L^{\rm free}$.  Explicitly
\be
\bar m_A^2 = g^2 \phi_0^2,  \qquad
\bar m_\eta^2=\bar m_\xi^2 = \xi g^2 \phi_0^2, \qquad 
\bar m_h^2 = V_{hh}, \qquad
\bar m_\theta^2 =V_{\theta\theta} +\bar m_\xi^2
\label{barmass}.
\ee
The time-independent frequencies are defined as before $\bar
\omega_i^2 = k^2 + \bar m_i^2$.  In real space
\ba
\bar G^{++}_{i}(x^\mu - y^\mu)  &=& 
\langle 0|T(\chi^{i}(x^\mu) \chi^{i}(y^\mu))|0\rangle =
\int \frac{\dd^4 k}{(2\pi)^4} \e^{-i k^\mu (x-y)_\mu} \bar G^{++}_{i}(k) 
\nn \\
&=&  \int \frac{\dd^3 k}{(2\pi)^3} \frac1{2\bar\omega_i}
\e^{-i k^\mu (x-y)_\mu} \Theta(x^0 - y^0) 
+ \int \frac{\dd^3 k}{(2\pi)^3} \frac1{2\bar \omega_i}
\e^{i k^\mu (x-y)_\mu} \Theta(y^0 - x^0)  \nn
\nn \\
&=& \bar G_{i}^{-+}(x^\mu-y^\mu) \Theta(x^0 - y^0) 
+ \bar G_{i}^{+-}(x^\mu-y^\mu)\Theta(y^0 - x^0)
\label{G_i}
\ea
and $\bar G_{i}^{--}(x^\mu - y^\mu) = \bar G_{i}^{++}(y^\mu -
x^\mu)$. In the second step we performed the contour integral over $k^0$.
A similar derivation can be done for the gauge boson propagators.  In
the one-loop calculation we only need certain contracted expressions.
These can be expressed in terms of the scalar propagator above
\eref{G_i}, with now $i= \{A,\xi\}$  (the equations
apply equally well to all ($\pm\pm$)-Green's functions).
\ba
g^\mn \bar G_{A^\mu A^\nu} &=& -3 \bar G_{A} -\xi \bar G_{\xi}
\label{Gtr} \\
g^{\mu\nu}  g^{\rho\sigma}  \bar G_{A^\nu A^\rho}\bar G_{A^\sigma A^\mu}&=&
 3 (\bar G_{A})^2 + \xi^2(\bar G_{\xi})^2
\label{Gsqr} \\
\bar G_{A^0A^0} &=& -(1-\bar \omega_A^2/\bar m_A^2) \bar G_A
-\xi (\bar \omega_\xi^2/\bar m_\xi^2) \bar G_\xi.
\label{G00}
\ea
The first expression is needed for the first order result (the gauge
boson loop), the second and third for the second order result (the gauge
boson loop and the mixed gauge boson-Goldstone boson loop
respectively).

For the one-loop calculation we only need the $+$-branch equal time
propagator:
\be
\bar G_i^{++}(0) = \int \frac{\dd^3 k}{(2\pi)^3} \frac1{2\bar\omega_i},
\label{G0}
\ee
where we used $\Theta(0) =1/2$.

\subsection{Dressed propagators} \label{dressed}

We will define dressed or resummed propagators, needed for the
non-perturbative one-loop correction. The quadratic part of the
potential, which has pieces in both $\L^{\rm free}$ and $\L^{\rm
  int}$, can be written in the form $\L^{\rm quad}=-(1/2)\sum_{i,j}
\chi_i(x^\mu) K^{ij}(x^\mu) \chi_j(x^\mu)$.  The dressed Green's
functions are defined as for the free case \eref{defG}, but now with
possible time-dependent pieces in the wave operator $K^{ij}$.  For the
one-loop calculation we only need the $(++)$-propagator, which we
discuss below; for ease of notation we drop the $(++)$-subscript.

The dressed Green's function satisfies the equation $K^{ij}(x^\mu)
G_{jk}(x^\mu -y^\mu) = -i\delta(x^\mu - \nolinebreak y^\mu) \delta_{ik}$.  Fields
with diagonal quadratic terms $K^{ij} \propto \delta^{ij}$ decouple
from the other fields, and we can express the Green's function in
terms of the mode functions in the usual way. For coupled fields, as
is the case with $A^0$ and $\theta$ in our case, something similar is
possible, but this involves more work.  Consider a real scalar with
canonical kinetic terms, then $ K^{ii} = \Box + m_i^2$.  Expand the
field
\be
\phi_i(x^\mu) = \int \frac{\dd^3 k}{(2\pi^3)} 
\frac{1}{\sqrt{2\bar\omega_{\vec k,i}}}
\[ a_{\vec k} U_{\vec k,i}(t) \e^{i \vec k \cdot \vec x}+
 a_{\vec k}^\dagger U^*_{\vec k,i}(t) \e^{-i \vec k \cdot \vec x} \],
\ee
with boundary conditions $U_{\vec k,i}(0) = 1, \; \dot U_{\vec k,i}(0)
= - i\bar\omega_{\vec k,i}$ such that $U_{\vec k,i}$ is the positive
frequency mode for scalar $\phi^i$. The Fourier transform of the Green's
function $G = \langle T(\phi(x^\mu)\phi({x'}^\mu))\rangle$ can then be
written in terms of the mode functions:
\be
G_{\vec k,i}(t,t') = \frac{1}{2\bar \omega_{\vec k,i}} \( U_{\vec k,i}(t) 
U_{\vec k,i}^*(t') \Theta(t-t')+
U_{\vec k,i}(t') U_{\vec k,i}^*(t) \Theta(t'-t) \).
\ee
The mode functions satisfy the wave equation with a time-dependent
frequency:
\be 
K^{ii}(t,\vec k)U_{\vec k,i}(t) = 
\[\partial_t^2 + \omega^2_i(t) \] U_{\vec k,i}(t) = 0,
\ee
such that $G_i(x-x') = \int\frac{\dd^3k}{(2\pi)^3} G_{\vec
  k,i}(t,t')$ indeed satisfies the Green's function equation.  To show
this use that the Wronskian $\dot U_{\vec k,i} U_{\vec k,i}^*- U_{\vec
  k,i} \dot U_{\vec k,i}^* = -2i\bar\omega_{\vec k,i}$ is constant in
time.

For the
one-loop calculation we only need the equal-time propagator which is
\be
G_i(0) =\int\frac{\dd^3k}{(2\pi)^3} \frac{|U_{\vec k,i}|^2}{2\bar \omega_{\vec k,i}} .
\ee
%

\subsection{Fermions}

First we go to a field basis where the mass matrix is diagonal. For
each fermionic field $\psi$ the quadratic part of the Lagrangian can
then be written as
\be
\L^{(2)}_\psi= \bar \psi K \psi
= \bar \psi \[ i\gamma^\mu \partial_\mu - m_\psi \] \psi .
\ee
%
The dressed propagator is defined as $K(x) D_\psi(x-y) = i \delta(x-y)
\mathbf{I}$, it is a Green's function of the Dirac operator.  As usual
we can expand the fermion field
\be
\psi= \sum_s \int \frac{\dd^3 k}{(2\pi)^3 \sqrt{2\bar \omega_{\vec{k}}}}
\[ b_{\vec k,s} u_{\vec k,s} \e^{i \vec k \cdot \vec x} 
+  d^\dagger_{\vec k,s} v_{\vec k,s} \e^{-i \vec k \cdot \vec x} \],
\ee
with $\{b_{\vec k,s}, b^\dagger_{\vec k',s'}\}= \{d_{\vec
  k,s},d^\dagger_{\vec k',s'}\} =(2\pi)^3 \delta(\vec k - \vec k')
\delta_{ss'}$.  For a Majorana spinor we have  $d_{\vec{k}} = \nolinebreak b_{\vec{k}}$, i.e. a particle is its
own anti-particle. The spinor function $u_{\vec k,s}$ satisfies the
equation $(i\partial_t -\nolinebreak \H_{\vec k}) u_{\vec k,s} = 0$ with $\H_{\vec
  k} =\gamma^0(\gamma^i k_i + m_\psi)$, the Fourier transformed
Hamiltonian.  Now we make the Ansatz
\be
u_{\vec k,s} = N \[i\partial_t + \H_{\vec k}\] U_{\psi}( \vec k) R_{s,u},
\qquad
v_{\vec k,s} = N \[i\partial_t + \H_{-\vec k} \] V_{\psi}(\vec k) R_{s,v}.
\label{Ansatz}
\ee
The spinors $R_s$ are helicity eigenstates, normalized such that
$R_s^\dagger R_{s'} =\delta_{s s'}$. Further  $\gamma^0 R_{s,u} =  R_{s,u}$ and
$\gamma^0 R_{s,v} = -R_{s,v}$.  The mode functions are each other's complex
conjugates: $V^*_{\vec k} = U_{\vec k}$.  Using usual free field
normalization for the mode functions at $t = 0$ gives $N =
1/\sqrt{\bar \omega_{\vec{k}} + \bar m_\psi}$ for the normalization factor.
The mode function equation is
\be
[\partial_t^2 + k^2 +m_\psi^2 -i \dot m_\psi  ] U_{\psi} =0.
\label{mode_psi}
\ee
This is of the same form as the mode equation for the scalar field,
namely a wave equation with time dependent frequency. Splitting the
frequency in a time-independent and dependent part gives $\bar
\omega_\psi^2 = k^2 + \bar m_\psi^2$ and $\delta \omega_\psi^2 =
\delta m_\psi^2 -i \dot m_\psi$.  It can be solved analogously to the
scalar field case.  Make the Ansatz
\be
U_\psi  = \e^{-i \bar \omega_\psi t} ( 1+ f_\psi), \qquad U_\psi(0) =1, 
\;\; \dot U_\psi(0) = -i \bar \omega_\psi.
\ee
The dressed equal-time propagator is now
\ba
G_\psi(0) &=& \langle  \psi(t) \bar \psi(t) \rangle =
\sum_s \int \frac{\dd^3k}{(2\pi^3) 2\bar \omega_\psi}  
 u_{\vec k,s} \bar u_{\vec k,s}
\nn \\
&=&
 \sum_{\rm d.o.f.} \frac12 \int \frac{\dd^3k}{(2\pi)^3}
\[ 1 - \frac{\bar \omega_\psi-\bar m_\psi}{\bar \omega_\psi} |U_{\psi}|^2 \]
\label{G_psi}.
\ea
The sum over the d.o.f. gives a factor 4 for a Dirac fermion, and a
factor 2 for a Majorana/Weyl fermion.  

\section{Perturbative calculation} \label{pert}

In this appendix we calculate the one-loop corrected equations of
motion in arbitrary gauge to show explicitly that the results are
gauge independent on-shell. The equations of motion in the CTP
formalism are given by $\delta \Gamma/\delta \phi^+|_{J_\pm=0} = 0$.
Diagrammatically this corresponds to all one-loop diagrams with one
external $h^+$ leg.  To isolate the divergent parts we need to go to
second order in coupling.

\subsection{First order}

At zeroth order the tadpole diagram contributes, and we recover the
classical equations of motion: $0 = i \Gamma^{+}_h = i(-i(\Box \phi +
V_\phi))$, see \eref{vertex}.  We write
\be
0=\Box \phi + V_\phi + A_1 + A_2 +{\rm finite}
\ee
with $A_1,A_2$ the first and second order contribution respectively
(with one and two vertex insertions respectively).  At first order four
diagrams contribute, with $\{h,\theta,\eta,A^\mu\}$ running in the
loop, and 
\ba
A_1 &=&\frac{i}2 \Gamma^+_{h \alpha\alpha} \bar G^{++}_\alpha(0)
\label{eom1} \\ &=&
 \frac12 \[ \left(\partial_\phi\delta m_h^2\right) \bar G_h^{++}(0) +
\left( \partial_\phi \delta m_\theta^2\right) \bar G_\theta^{++}(0)
-2 \left(\partial_\phi \delta m_\eta^2\right) \bar G_\eta^{++}(0)
+ \left(\partial_\phi \delta m_{A^\mu A^\nu}^2\right) \bar G_{A^\mu A^\nu}^{++}(0) \] .
\nn
\ea
The overall half factor is a symmetry factor, relating to the
reflection symmetry of the Feynman diagrams.  The two-point vertex is $
\Gamma^{+}_{ii} =- i \delta m^2_{ii} $.  The $\eta$ loop picks up a
minus sign because of the anti-commuting nature of $\eta$, and a
factor two for the two fermionic degrees of freedom.  The gauge boson
term can be rewritten using \eref{Gtr}:
\be
(\partial_\phi \delta m_{A^\mu A^\nu}^2) \bar G_{A^\mu A^\nu}^{++}(0) 
= (- \partial_\phi\delta m_A^2) g^\mn \bar G_{A^\mu A^\nu}^{++}(0)
=(\partial_\phi\delta m_A^2) (3 \bar G_A^{++}(0) + \xi \bar G_\xi^{++}(0)).
\ee
Taking the large momentum limit, the equal time propagator \eref{G0} behaves as
\be
\bar G^{++}_i(0) 
= \frac1{4\pi^2} \int k^2\dd k  \[ \frac{1}{k} -\frac12 \frac{\bar m_i^2}{k^3} +...\]
= \frac1{8\pi^2} \[\Lambda^2 -\frac12 \bar m_i^2 \ln \Lambda^2 + {\rm finite}\].
\ee
Thus $A_1$ becomes \eref{eom1}:
\ba
A_1 &=& \frac{\partial_\phi}{16\pi^2} 
\[\delta m_h^2 + \delta m_\theta^2 -2 \delta m_\eta^2  +3 \delta m_A^2
+ \delta m_\xi^2\]  \Lambda^2
\label{Vfirst}\\ &&
-\frac12
\[\partial_\phi \delta m_h^2 \bar m_h^2+\partial_\phi \delta  m^2_\theta \bar m^2_\theta - 2
\partial_\phi\delta m^2_\eta \bar m^2_\eta
+3 \partial_\phi \delta m_A^2 \bar m_A^2 + \partial_\phi\delta
m_\xi^2 \bar m_\xi^2\] 
\ln \Lambda^2.
\nn
\ea
Here we defined $m_\xi^2 = \xi m_A^2$ analogous to \eref{barmass}.
Upon inserting explicit mass terms, we infer that the quadratic divergence is gauge
independent, but that the log-divergence depends on $\xi$.  As we will see,
this gauge dependence is cancelled by the second order term.

\subsection{Second order}

Consider first the diagonal loops with a single field running in the
loop, and a three and two-point vertex insertion.  The mixed loop,
with propagators for both $\theta$ and $A^0$ is discussed afterwards.
For each field running in the loop there are two diagrams that
contribute, one with a $\Gamma_{\alpha\alpha}^+$ mass insertion and
$G^{++}_\alpha$ propagators, and one with a $\Gamma_{\alpha\alpha}^-$
mass insertion and $G^{+-}_\alpha$ propagators.  Let us start with the
Higgs boson loop $h$.  Its contribution to the equations of motion at
second order is
\ba 
A_2  \! &\supset& \! \frac{i}2
\int \dd^4 x' \Gamma_{hhh}^{+}(x) \bigg[
\bar G_h^{++}(x\!-\!x') \Gamma^{+}_{hh}(x') \bar G_h^{++}(x'\!-\! x)
+\bar G_h^{+-}(x\! -\! x') \Gamma^{-}_{hh}(x') \bar G_h^{-+}(x'\! -\!x)
\bigg]
\nn\\
&=&-\frac{i}2 (\partial_\phi\delta m^2_h(t)) \int \dd^4 x'  \delta m^2_h(t')
\[ \bar G_h^{++}(x-x')^2 - \bar G_h^{+-}(x-x')^2\].
\ea
Here we used for the two-point function $\Gamma_{hh}^{+} = - \Gamma_{hh}^{-} =- i \delta
m^2_h$. The overall symmetry factor $1/2$ originates, again, from a reflection
symmetry.  Plugging in the expressions for the propagators gives
\ba
&&\int \dd^4 x' \[\bar G_h^{++}(x-x')^2 - \bar G_h^{+-}(x-x')^2\] 
\nn \\ && =
\int \dd^4  x' \!\! \int\frac{\dd^3 k}{(2\pi)^3 2\omega_k} 
\frac{\dd^3 p}{(2\pi)^3 2\omega_p}
\!\[\e^{-i( k + p)(x-x') }\Theta(t\!-\!t') +\e^{-i( k +
  p)(x'-x)} \Theta(t'\!-\!t)
-\e^{i( k + p)(x-x')} \]
\nn \\
&& =\int \dd t' \int \frac{\dd^3 k}{(2\pi)^3}
\frac{-2i}{(2\bar{\omega}_{\vec k,h})^2}
\[\sin[2\bar{\omega}_{\vec k,h} (t-t')]\Theta(t\!-\!t')  \] .
\ea
To get to the second line, we used that integration over $\vec x'$ gives
a factor $\delta^3(\vec k + \vec p)$.  This can be integrated over
$\vec p$, which sets $\bar{\omega}_{\vec{k}} =
\bar{\omega}_{\vec{p}}$.  Putting it all together the $h$-loop
contributes 
\be
A_2 
\! \supset \!   -\int \frac{\dd^3 k}{(2\pi)^3(2\bar{\omega}_{\vec k,h})^2}
 \delta m^2_h(t)\int \dd t'  \delta m^2_h(t')
\[\sin[2\bar{\omega}_{\vec k,h} (t-t')]\Theta(t\!-\!t') \].
\ee
To extract the divergent part we partially integrate:
\ba
\int \dd t' \delta m^2(t') \sin(2\omega(t\!-\!t')) \Theta(t\!-\!t') 
&=& \frac{ \delta m^2(t')}{2\omega} \cos(2\omega(t\!-\!t'))\bigg|^{t'=t}_{t'=t_0}
-\!\int \dd t'  \frac{\delta\dot m^2(t') }{2\omega} \cos(2\omega(t\!-\!t'))
\nn \\ &=& \frac{ \delta m^2(t')}{2\omega} + \O(\omega^{-2}),
\ea
where we set $\delta m^2(t_0) =0$ at the initial time.   And thus
\be
A_2 \supset -(\partial_\phi \delta m^2_h(t))  \delta m^2_h(t) \int \frac{\dd^3 k}{(2\pi)^3}
\frac{1}{(2\bar{\omega}_{\vec k,h})^3} + \O(\bar{\omega}_{\vec k,h}^{-4})
= -\frac{1}{32\pi^2}  (\partial_\phi \delta m^2_h)  \delta m^2_h
\ln \Lambda^2 + {\rm finite}.
\label{V2nd}
\ee
In the last step we expanded in large $|\vec k|$.

The calculation of the $\theta$ and $\eta$ loops proceeds analogously, and
gives a contribution just as \eref{V2nd} with the appropriate mass; in
addition the $\eta$-loops picks up an overall factor $(-2)$ because of
the two anti-commuting d.o.f.  The contribution for the gauge field is
\ba 
A_2 &\supset&
-\frac{i}2  \partial_\phi \delta m^2_A(t)\int \dd^4 x' \delta m^2_A(t') 
g^{\mu \rho} g^{\nu \sigma}
\[ \bar G_{A^\mu A^\nu}^{++} \bar G_{A^\rho A^\sigma}^{++}
 - \bar G_{A^\mu A^\nu}^{+-} \bar G_{A^\rho A^\sigma}^{-+}\]
\nn\\
 &=&
-\frac{i}2 \partial_\phi \delta m^2_A(t)\int \dd^4 x' \delta m^2_A(t') 
\[ 3\( (\bar G_{A}^{++})^2 - (\bar G_{A}^{+-})^2\)
 +\xi^2\((\bar G_{\xi}^{++})^2 - (\bar G_{\xi}^{+-})^2\)
\]
\nn \\
&=& -\frac{1}{32\pi^2} \[3 (\partial_\phi\delta m^2_A)\delta m^2_A
+ (\partial_\phi\delta m^2_\xi) \delta m^2_\xi\]\ln \Lambda^2 + {\rm finite}.
\ea
In the first line we used the definition of mass \eref{mass} $m_{A^\mu
  A^\nu} = -g^\mn m_A^2$ with $m_A^2 = g^2 \phi^2$. To get the second
line we used \eref{Gsqr}.  The expression has been reduced to a sum of
two scalar integrals, which result in expressions analogous to \eref{V2nd} to give 
the final result, given in the last line above.  Adding it all up gives
\ba
A_2^{\rm diag} =
- \frac{1}{32\pi^2}  \bigl[ &&(\partial_\phi\delta m_h^2) \delta m_h^2
+ (\partial_\phi \delta m_\theta^2)  \delta m_\theta^2
-2 (\partial_\phi\delta m_\eta^2)  \delta m_\eta^2+ 3 (\partial_\phi\delta m^2_A)
 \delta m^2_A\nn\\
 &&\qquad+(\partial_\phi \delta m^2_\xi)  \delta m^2_\xi \bigr]\ln \Lambda^2  .
\ea

In $\L^{\rm int}$ there is also a derivative interaction mixing the
gauge and the Goldstone boson. This leads to a mixed loop diagram.
Since $\phi(t)$ does not depend on spatial coordinates, the
derivatives will only act on time, and thus the mass terms contain
factors $g^{00}$.  The mixed diagram contributes
\ba 
A_2^{\rm mix} &=&
-{i} \partial_\phi\delta m^2_{A\theta} (t)\int \dd^4 x' \delta m^2_{A\theta}(t') 
g^{0\mu} g^{0 \nu}
\[ \bar G_{A^\mu A^\nu}^{++} \bar G_\theta^{++}
 - \bar G_{A^\mu A^\nu}^{+-} \bar G_\theta^{-+}\]
\\
&=&
i  \partial_\phi\delta m^2_{A\theta} (t)\int \dd^4 x' \delta m^2_{A\theta}(t') 
\[ \(1- \frac{\bar \omega_A^2}{\bar m_A^2}\) \bar G^{++}_A
+ \frac{\xi\bar \omega_\xi^2}{\bar m_\xi^2 } G_\xi^{++} \] G_\theta^{++}  
-(++ \to +-) .\nn
\ea
There is no symmetry factor $1/2$ since there is no reflection
symmetry.  In the first line we used $\Gamma^{+}_{A^\mu \theta} = i
\delta m_{A\theta}^2 g^{0\mu}$, and $m_{A\theta}^2= \delta
m_{A\theta}^2$.  Using \eref{G00} we reduced the propagators to scalar
propagators as before.  Plugging in the explicit expressions we find
\ba
A_2^{\rm mix} &=& 2
\partial_\phi \delta m^2_{A\theta} (t)\int \dd t' \delta m^2_{A\theta}(t') 
\int \frac{\dd^3 k}{(2\pi)^3}
\bigg\{\Big[\frac{1-\bar \omega_A^2/\bar m_A^2}{4\bar \omega_\theta \bar \omega_A} 
\sin((\bar \omega_A + \bar \omega_\theta)\Delta t)
\nn \\ && +\frac{\xi \bar \omega_\xi^2/\bar m_\xi^2 }{4\bar \omega_\theta \bar \omega_\xi} 
\sin((\bar \omega_\xi + \bar \omega_\theta)\Delta t)
\Big] \Theta(\Delta t) \bigg\}
\nn\\
 &=&
2
(\partial_\phi\delta m^2_{A\theta} ) \delta m^2_{A\theta} \int \frac{\dd^3 k}{(2\pi)^3}
\[\frac{1- \bar \omega_A^2/ \bar m_A^2}
{4\bar \omega_\theta \bar \omega_A(\bar \omega_\theta+ \bar \omega_A)} 
+\frac{\xi \bar \omega_\xi^2/\bar m_\xi^2 }
{4\bar \omega_\theta \bar \omega_\xi(\bar\omega_\theta+ \bar \omega_\xi)} 
\]+ \O(\omega_i^{-4})
\nn\\
&=&
\frac{(3+\xi)}{64\pi^2} (\partial_\phi\delta m^2_{A\theta} ) \delta m^2_{A\theta}\ln \Lambda^2 +{\rm finite}.
\ea
In the second step we performed a partial integration to isolate the
divergent parts.

The one-loop correction to the equation of motion is $A_1 + A_2^{\rm
  diag} + A_2^{\rm mix}$, which gives
\ba
0&=&\Box \phi + V_\phi + \frac{\partial_\phi}{16\pi^2} 
\[\delta m_h^2 + \delta m_\theta^2 -2 \delta m_\eta^2  +3 \delta m_A^2
+ \delta m_\xi^2\]  \Lambda^2
\nn\\
&-&\frac{1}{32\pi^2}
\[(\partial_\phi  m_h^2)m_h^2+(\partial_\phi m^2_\theta) m^2_\theta - 2
(\partial_\phi m^2_\eta)  m^2_\eta
+3 (\partial_\phi  m_A^2) m_A^2 + (\partial_\phi
m_\xi^2)  m_\xi^2\] 
\ln \Lambda^2 \nn\\
&+&\frac{(3+\xi)}{64\pi^2} (\partial_\phi m^2_{A\theta} ) m^2_{A\theta} \ln
\Lambda^2
\ea
where we used $\partial_\phi \delta m^2_\alpha = \partial_\phi m^2_\alpha $.

Integrating to get the effective action gives:
\ba
\Gamma^{1-{\rm loop}} &=& - \frac{1}{16\pi^2} \int \dd^4 x \bigg\{
\[m_h^2 + m_\theta^2 -2m_\eta^2  +3m_A^2 + m_\xi^2\] \Lambda^2
\\ 
&& \hspace{2cm}-\frac14
\[m_h^4+  m^4_\theta - 2 m^4_\eta
+3 m_A^4 + m_\xi^4 -\frac12(3+\xi) m_{\theta A}^4\] \ln \Lambda^2 \bigg\}
\label{V}
\nn\\
&=&
-\frac{1}{16\pi^2} \int \dd^4 x  \bigg[ \Lambda^2 \(V_{hh}+V_{\theta\theta}+3 m_A^2\)
-  \frac{\ln \Lambda^2}{4} \(V_{hh}^2+V_{\theta\theta}^2+3 m_A^4
-6  V_{\theta \theta} m_A^2 \) \bigg].\nn
\ea
plus finite and field-independent terms.   The gauge parameter $\xi$ cancels, and gauge invariance of the final
 result is manifest,  provided we use the zeroth order equation of
 motion (together with gauge invariance) to write $\delta m^4_{A\theta} =
4 m_A^2 V_{\theta\theta}$ \eref{eom_zero}. 
To get the final
result we inserted the explicit form of the masses from \eref{mass},
and the definition $m_\xi^2 = \xi m_A^2$.  This result is in agreement
with the non-perturbative calculation \eref{result_np}. The gauge dependent part of
$m_\theta^2$ and $m_A^2$ cancels against that of the ghosts, i.e.
$(m_\theta^2 +m_\xi^2 - 2 m_\eta^2) =V_{\theta\theta}$, making the
quadratic terms coming from the first order calculation gauge invariant.
Combining the first and second order calculation renders also the logarithmic
divergences gauge invariant, but only upon using the equations of
motion \eref{eom_zero}: $(m_\theta^4-2m_\eta^4 +m_\xi^4 -2\xi
V_{\theta \theta} m_A^2) = (V_{\theta\theta})^2$.

\end{document}